\documentclass[english,aps,prb,floatfix,twocolumn,showpacs,amsmath,amssymb,eqsecnum,superscriptaddress]{revtex4}
\usepackage[T2A,T1]{fontenc}
\usepackage[latin9]{inputenc}
\usepackage{color}
\definecolor{note_fontcolor}{rgb}{0.800781, 0.800781, 0.800781}
\usepackage{babel}
\usepackage{textcomp}
\usepackage{amsmath}
\usepackage{graphicx}
\usepackage[unicode=true,pdfusetitle,
 bookmarks=true,bookmarksnumbered=false,bookmarksopen=false,
 breaklinks=false,pdfborder={0 0 1},backref=false,colorlinks=false]
 {hyperref}

\makeatletter

\providecommand{\tabularnewline}{\\}

\@ifundefined{textcolor}{}
{%
 \definecolor{BLACK}{gray}{0}
 \definecolor{WHITE}{gray}{1}
 \definecolor{RED}{rgb}{1,0,0}
 \definecolor{GREEN}{rgb}{0,1,0}
 \definecolor{BLUE}{rgb}{0,0,1}
 \definecolor{CYAN}{cmyk}{1,0,0,0}
 \definecolor{MAGENTA}{cmyk}{0,1,0,0}
 \definecolor{YELLOW}{cmyk}{0,0,1,0}
}

\newcommand{\angstrom}{\text{\normalfont\AA}}
\usepackage{hyperref}
\usepackage{textcomp}

\@ifundefined{showcaptionsetup}{}{%
 \PassOptionsToPackage{caption=false}{subfig}}
\usepackage{subfig}
\makeatother

\begin{document}
\preprint{This line only printed with preprint option}
\title{Tuning topological surface states by cleavage angle in topological
crystalline insulators}

\author{Evgeny Plekhanov}
\email{evgeny.plekhanov@kcl.ac.uk}

\affiliation{King's College London, Theory and Simulation of Condensed Matter (TSCM),
The Strand, London WC2R 2LS, United Kingdom}
\author{Cedric Weber}
\email{cedric.weber@kcl.ac.uk}

\affiliation{King's College London, Theory and Simulation of Condensed Matter (TSCM),
The Strand, London WC2R 2LS, United Kingdom}
\begin{abstract}
The conducting states, recently discovered at the surface of two special
class of insulators -- topological insulators and topological crystalline
insulators - are distinguished by their insensitivity to local and
non-magnetic surface defects
at a level of disorder, sufficiently small to be described within the perturbation theory.
However, the behavior of the surface states in case of non local macroscopic imperfections
is not clear.
Here, we
propose a systematic study of the topological surface states on vicinal planes (deviations
from perfect surface cleavage) in a topological crystalline insulator
of the tin telluride family, by using realistic first-principles-derived
tight-binding models. The theoretical framework proposed is quite
general and easily permits the extensions to other topological insulator families.
\end{abstract}
\pacs{71.20.\textminus b, 71.70.Ej, 73.20.\textminus r, 73.20.At}
\maketitle

\section{introduction}

Topological surface states (TSS), occurring in topological insulators
(TI), are, probably, the most exciting and exotic discoveries in condensed
matter physics in the last years\citep{TI,Qi_2011}. Experimental
realization of the (TI)\citep{xia_2009,zhang_2009} paves the way
to numerous potential technological applications: the quantum spin
Hall effect, the dissipationless spin current, the magnetoelectric
effect \citep{KaneMele_2005,Bernevig_2006,Xu_2006,Qi_2008} etc. The
discovery of TSS has opened a race to search for topological states
protected by some other symmetries. Along this line, the theoretical
proposal of Fu about the TSS protected by crystalline symmetry
\citep{TCI,Slager_2012,Slager_2015,Slager_2017,Slager_2018}
has soon found experimental confirmation \citep{SnTe3,SnTe1,SnTe2}
in tin telluride (IV-VI) family and stimulated intense work on the
search for materials in the new class of topological crystalline insulators
(TCI). At variance with the conventional TI, where the protection
of the TSS comes from time-reversal symmetry, in TCI the protection
is ensured by the crystalline symmetry, usually the mirror symmetry.
TCI are characterized by a new topological invariant - the so-called
mirror Chern number $nM$ - analogously to the TI, which are characterized
by the $Z_{2}$ topological invariant. A system can be a trivial TI,
but possess a non-zero $nM$, which is exactly the case of tin telluride
SnTe -- a prototypical TCI. For the firm observation of topological
states in experiments it is crucial to know how the defects of various
types, always present in real materials, could influence the TSS.
Thanks to the topological protection, the TSS should be quite robust
against the local (non-magnetic) surface defects. Moreover, it was
demonstrated recently that the TSS are also insensitive to the disorder
in the bulk \citep{di_sante_2015}. 

A different type of defect can be described as a slight deviation
of the surface cut from the most common and highly symmetric one.
Indeed, the particularity of the TSS lies in the fact that they only
appear on the particular cuts of the bulk crystal (e.g. in SnTe, the
topological surfaces are $(001)$, $(111)$ and $(110)$ \emph{c.f.}
Ref.\onlinecite{buczko}). In the real experiment, the surface might
not be cleaved precisely at the right angle and, therefore, a legitimate
question arises: to which extent the crystal surface can deviate from
the ideal one so that the TSS are still present? %
For completeness, we note that in the case of $Z_{2}$ topological insulators the answer is quite clear:
strong topological insulators will have SS on every surface, while
weak ones only on some of them depending on how the 
time-reversal symmetry (TRS) points are
projected on the surface~\cite{Fu_Kane_2007}, while no conclusion a priori can be made for TCI.
It was pointed out in Ref.~\onlinecite{SnTe1} that TSS arise in SnTe if the projections of the TRS points on
the surface possess the mirror symmetry. Thus, one needs not only to trace the projections of the TRS
points, but also to make sure that these projections lie within a mirror plane.

To answer the above question, realistic, material-specific calculations
of large systems are needed. The \emph{ab-initio} methods can easily reach
their computational limits due to the mandatory use of the spin-orbit
coupling in the simulations and the need to well exceed the critical
slab thickness in order to observe the TSS. Thus, one resorts to the
realistic tight-binding models with the parameters chosen so as to
reproduce the \emph{ab-initio} band-structure. In this way, large
super-cell calculations can be easily afforded at low computational
cost, while conserving the predictive power of \emph{ab-initio} approaches.
In such a study, the tin telluride family (SnTe, Pb$_{1-x}$Sn$_{x}$Se,
and Pb$_{1-x}$Sn$_{x}$Te) represents almost an ideal playground
thanks to the simplicity of the unit cell and the richness of the
phase diagram.

In SnTe and in other tellurides, vicinal planes represent a commonly occurring example of
a non-ideally cleaved surface~\cite{Deringer_2015,DiSante_2016,Zallo_2017}. 
In this article we study the TSS on vicinal surfaces, which deviate
from the ideal ones. To achieve this, we construct the super-cells
with the so-called tilted boundary conditions, so that one of the
boundaries of the super-cell appears to make a finite angle with respect
to the crystallographic axes of the unit cell.
As we will show below, in this case, there is always at least one mirror plane, which is perpendicular to the tilting
axis and to the tilted surface and which passes through some of the projections of TRS points, thus
ensuring the mirror symmetry in those projections.

In this setup, some of the TSS will have the topological protection, since they are projections of
bulk TRS and have a mirror plane,
although their precise positions and
the form of band dispersion is to be determined. We combine high performance slab calculations with
the TRS projection analysis and show how the topologically protected and unprotected states
evolve upon changing the cleavage
angle with respect to the three topological surfaces in SnTe.

This article is organized as follows: in Sec.\ref{sec:methods}, we
describe the computational methods used; in Sec.\ref{sec:results},
we present the numerical results, while further discussions and conclusions
are given in Sec.\ref{sec:Conclusions}. 

\section{\label{sec:methods}methods}

We perform Density functional theory (DFT) simulations using the Vienna
Ab initio Simulation Package (VASP) \citep{vasp} and the Generalized
Gradient Approximation (GGA)\citep{gga} in the Perdew-Burke-Ernzerhof
(PBE) formalism for the exchange-correlation potential. We use an
energy cutoff for the plane wave basis of $400$ eV and a $16\times16\times16$
Monkhorst-Pack $k-$point mesh\citep{mp}. Here we consider $[001]$,
$[110]$ and $[111]$ surfaces in the cubic phase. To calculate TSS
in the slab geometry, the number of layers has to be sufficiently
large to prevent the interaction between TSS of the two slab surfaces,
which makes \emph{ab-initio} approaches prohibitive. We, therefore,
resort to an effective tight-binding (TB) model. The TB hopping matrix
elements are determined by projection of the \emph{ab-initio}
VASP Hamiltonian onto the atomic-like orbitals through the WANNIER90
package\citep{wannier90}. In these projections we retain $s$- and
$p$-type basis functions. As for structural parameters, we employ
those optimized within DFT-GGA. For the cubic structure, we use lattice
constant $a=6.42$\angstrom, rhombohedral angle $\alpha=60^{\circ}$
in the unit cell with center of inversion (see Supplemental Materials
in Ref.\onlinecite{Evgeny_SnTe}.) 

In order to safely conclude about the presence of TSS, we plot the
2D band structure for the slabs with maximal possible thickness. This
essentially stands for many diagonalisations (as many as the number
of surface $k$-points) of rather large complex Hermitian matrices
(up to $30000\times30000$ in this work). We solve this technical
problem by employing parallel GPU diagonalization routines and CUDA/C/Fortran
interfaces.

We explain now the geometry conventions used in the present work in
order to define the tilted or vicinal planes. It is well known that
in SnTe the TSS are only present on three crystallographic surfaces.
These are $(001)$, $(111)$ and $(110)$\citep{buczko}. Each of
these cases corresponds to a unit cell, which possesses the corresponding
surface. Here we consider each of these unit cells $V$ with unit
vectors $\left\{ a_{1,},a_{2},a_{3}\right\} $ and build a new unit
cell $V^{\prime}$ with unit vectors \textbf{$\left\{ b_{1},b_{2},b_{3}\right\} $},
so that each $b_{i}$ is a linear combination of $\left\{ a_{j}\right\} $
with integer coefficients, so that $V^{\prime}$ always contain an
integer number of original cells $V$. We choose the original cells
to be orthorhombic and double if necessary the primitive cell. For
completeness we report below both the original cell $V$ and the tilted
one $V^{\prime}$ geometries.

\section{\label{sec:results}results}

\subsection{\label{subsec:Tilted001}Tilted (001) surfaces}

\begin{figure}
\includegraphics[angle=270,width=4.25cm]{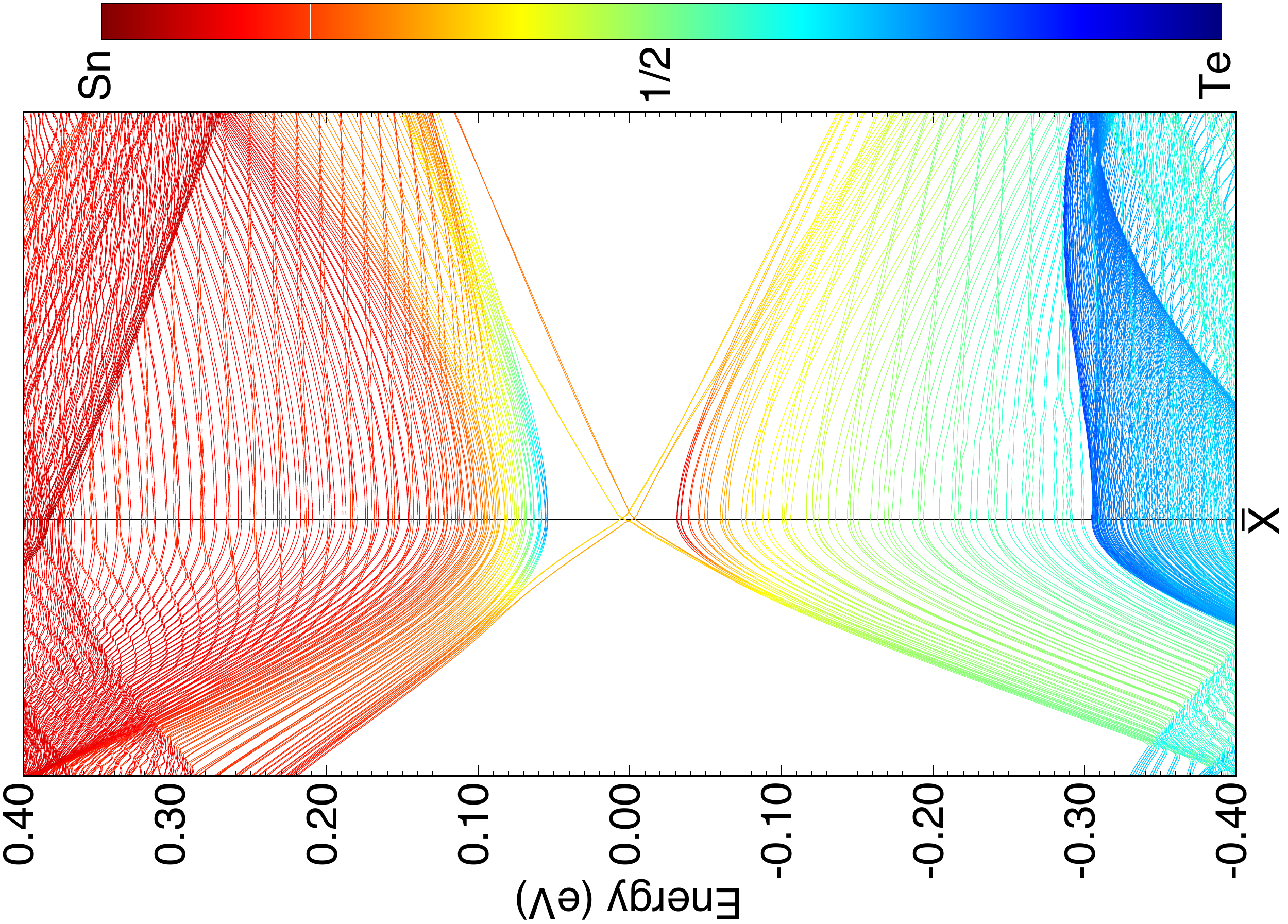}
\includegraphics[angle=270,width=4.25cm]{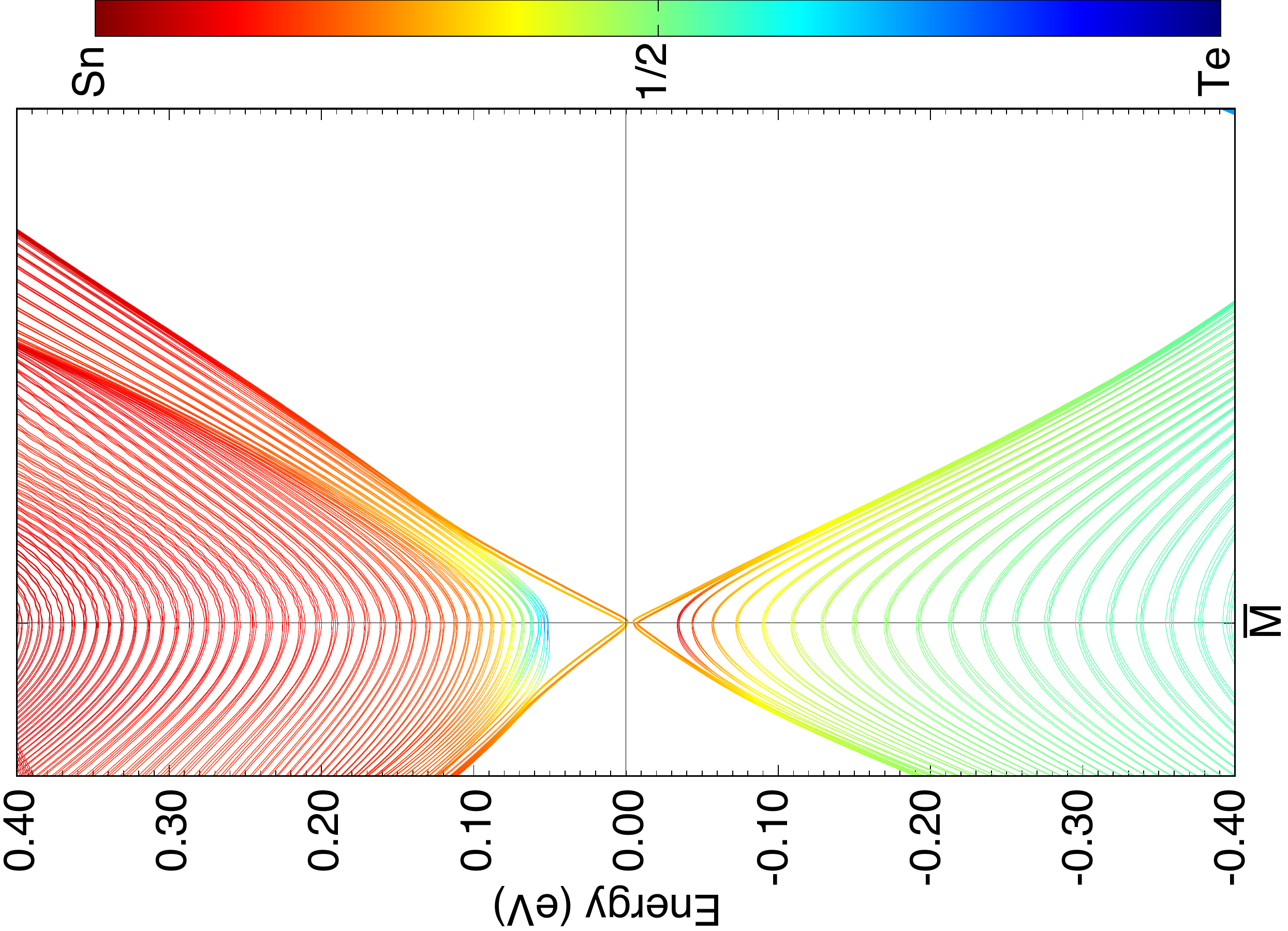}

\caption{\label{fig:TSS_001}(color online) 
	  Surface band structure for a slab with tilted $(001)$ surface
	  along the path $\overline{M}-\overline{X}-\overline{\Gamma}$
	  zoomed around $\overline{X}$ (left panel), and 
	  $\overline{Y}-\overline{M}-\overline{X}$ zoomed around $\overline{M}$ (right panel). The
	  line color reflects the orbital character of the bands: red - predominantly Sn, blue -
	  predominantly Te.
	  $n=3,$ $m=1,$ which corresponds to the angle of $\alpha=13.26^{\circ}$.
	  $30$ layers. For full details of the unit cell see Tab.\ref{tab:UC001}.
	  Note the topological protection in the left panel and the absence thereof in
	  the right one.
}
\end{figure}

The case of tilted $(001)$ surfaces is the most straightforward one.
We start from a tetragonal unit cell ($V$) which has the following
unit vectors (in Cartesian coordinates): $a_{1}=\frac{a}{2}(1,1,0),\ a_{2}=\frac{a}{2}(1,-1,0),\ a_{3}=a(0,0,1)$,
so that $a_{3}$ is perpendicular to $(001)$ surface. Here $a$ is
the SnTe lattice constant. The unit vectors $\{b_{i}\}$ of $m\times n$
$(001)$ tilted cell $V^{\prime}$ are expressed in Cartesian coordinates as follows:
\begin{eqnarray*}
b_{1} & = & a\left(\begin{array}{c}
n\\
n\\
m
\end{array}\right),\quad b_{2}=\frac{a}{2}\left(\begin{array}{c}
\phantom{-}1\\
-1\\
\phantom{-}0
\end{array}\right),\quad b_{3}=a\left(\begin{array}{c}
-m/2\\
-m/2\\
\phantom{-}n\phantom{/2}
\end{array}\right).
\end{eqnarray*}
It is easy to see that $b_{1}=2na_{1}+ma_{3}$, while $b_{3}=-ma_{1}+na_{3}$,
and $V^{\prime}$ is rotated by an angle $\vartheta$ around $b_{2}$
with respect to $V$. This cleavage angle $\vartheta$ can be found
as: $\tan\vartheta=m/\sqrt{2}n$. This choice of basis ensures the orthorhombicity
of the unit cell and allows to build up a sequence of surfaces with
the cleavage angle gradually approaching zero. 
Such a choice ensures minimal cluster size at a given angle $\vartheta$.
The values of $n$ and $m$ explored in the present work are listed in
Tab.\ref{tab:UC001}. 
\begin{figure*}
\begin{minipage}[t]{0.33\textwidth}%
\subfloat[$m=1$, $n=3$]{\includegraphics[viewport=120bp 80bp 900bp 910bp,clip,width=1\textwidth]{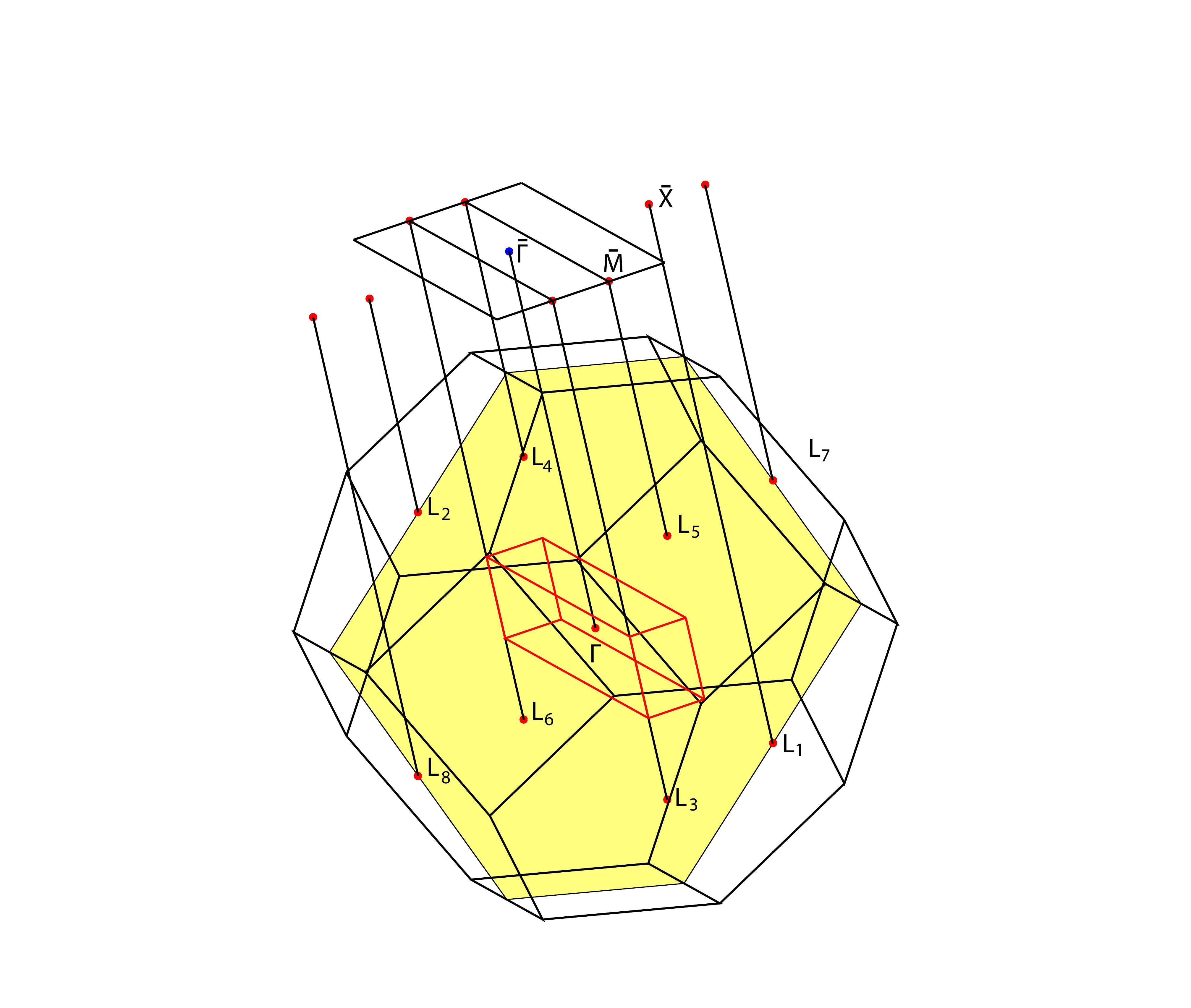}
}%
\end{minipage}\hfill{}%
\begin{minipage}[t]{0.33\textwidth}%
\subfloat[$m=2$, $n=4$]{\includegraphics[viewport=120bp 80bp 900bp 910bp,clip,width=1\textwidth]{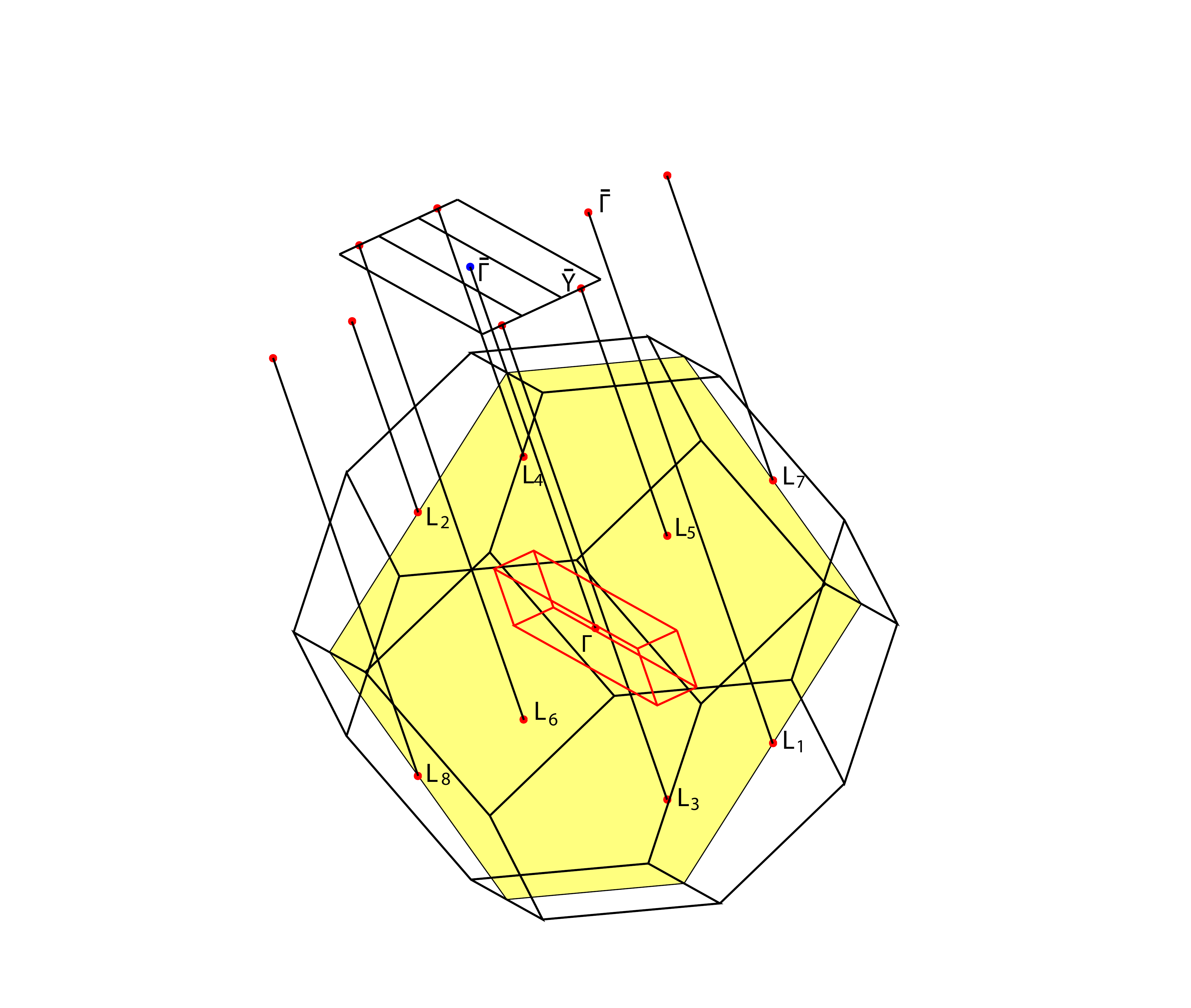}}%
\end{minipage}\hfill{}%
\begin{minipage}[t]{0.33\textwidth}%
\subfloat[$m=3$, $n=5$]{\includegraphics[viewport=120bp 80bp 900bp 910bp,clip,width=1\textwidth]{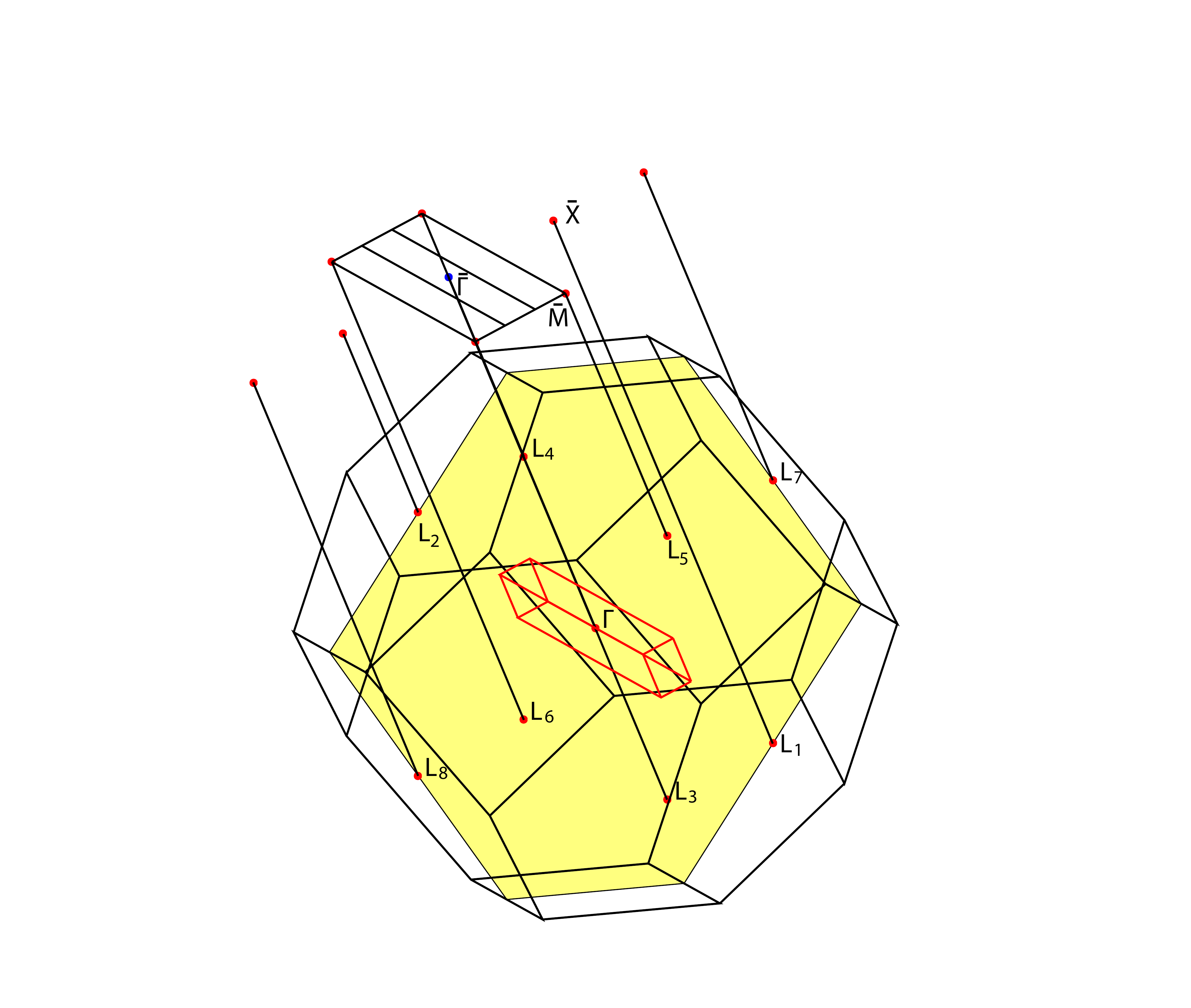}
}%
\end{minipage}

\caption{\label{fig:BZ_001}(color online)
	  The projections of $L$ TRS points of the rhombohedral Brillouin zone to a $(001)$ tilted
	  surface $3\times1$ (left panel); $4\times2$ (middle panel); $5\times3$ (left panel). The
	  corresponding folded BZs are also drawn in red. Note the projecting on $\overline{M}$ and
	  $\overline{X}$ for cases (a) and (c), while projecting on $\overline{Y}$ and
	  $\overline{\Gamma}$ for case (b). Shown in yellow is the $(1\bar{1}0)$ mirror plane.
}

\end{figure*}

At $\vartheta=0$ (the pristine $(001)$ surface), there are two mirror planes: $(1\bar{1}0)$ and $(110)$.
It it easy to see that at finite $\vartheta$ the mirror plane $(110)$ is lost, while the $(1\bar{1}0)$ is
preserved since the rotation axis $b_2$ is normal to it. Therefore, we conclude that the projections
of the TRS points on a tilted $(001)$ surface will have the topological protection if they lie
within the $(1\bar{1}0)$ mirror plane. We will show below that this is indeed the case and derive the
general rule describing the location of TSS for given $n$ and $m$.
An example of the slab dispersion containing TSS in the case of $n=3$,
$m=1$ is shown in Fig.~\ref{fig:TSS_001}.

The surface states appear at high-symmetry points of the surface Brillouin
zone depending on the parity of $m$ and independently on $n$. Namely, for even $m$, the surface
states are located at $\overline{\Gamma}$ and $\overline{Y}=\left(0,\pi\right)$, while
for odd $m$ they are at $\overline{X}=\left(\pi,0\right)$ and $\overline{M}=\left(\pi,\pi\right)$. This
alternation rule holds for all values of $m$ and can be understood
if one takes into account how $L$ points of the original rhombohedral
BZ are projected onto the surface BZ of $(001)$ tilted unit cell.
Nevertheless, the topological protection is only ensured for the $\overline{\Gamma}$ and 
$\overline{X}$ points as they are
crossed by the $(1\bar{1}0)$ mirror plane, while $\overline{Y}$ and $\overline{M}$ are not
protected by the mirror symmetry.
We report the 3D plots of such projections for $m=1,2,3$ in Fig.\ref{fig:BZ_001}.

It can be seen from Fig.~\ref{fig:TSS_001}, how the lack of topological protection changes the
low-energy physics. Namely, at $\overline{M}$, a tiny gap opens, while at $\overline{X}$ there are
topological states. The reason for this tiny gap is the numerical round-off errors inevitable in any
numerical calculation. Mirror symmetry at $\overline{X}$ make the TSS insensible to these errors,
while at $\overline{M}$ there is no such protection.
It is interesting to note that each of the eight $L$ points of the
first primitive BZ is projected onto a different $\overline{Y}$ (or
$\overline{M}$) point in different folded surface BZ.
Moreover, already at minimal $m=1$ the projections of some of the $L$ points in the first bulk BZ belong
to higher surface BZs. 
As $n\to\infty$ (and $\vartheta\to0$),
the projections move more and more towards higher surface BZs. 
On the other hand, as $n\to\infty$, the extension of the surface BZ
along $x$ direction tends to zero as well as the difference between
$\overline{Y}$ and $\overline{M}$ and between $\overline{\Gamma}$ and $\overline{X}$.
In addition, in this limit the $(110)$ mirror plane is restored. 
Therefore, in the limit $\vartheta=0$
each surface TRS point acquires the projections from two $L$ points,
which then form a bonding and anti-bonding combinations - a well known
fact for $(001)$ TSS in SnTe.\citep{SnTe1,Polley_2018}. Thus studying
$(001)$ tilted states allows to approach this limit gradually and
observe the progressive transformation of TSS.

\begin{table}
\begin{tabular}{|c|c|c|c|c|c|}
\hline 
$\vartheta,\;^{\circ}$ & $n$ & $m$ & $N_{states}$ & TP & TNP \tabularnewline
\hline 
\hline 
$23.00$ & $5$ & $3$ & $1888$ & $\overline{X}$ & $\overline{M}$ \tabularnewline
\hline 
$19.47$ & $4$ & $2$ & $1152$ & $\overline{\Gamma}$ & $\overline{Y}$ \tabularnewline
\hline 
$13.26$ & $3$ & $1$ & $608$  & $\overline{X}$ & $\overline{M}$ \tabularnewline
\hline 
$10.03$ & $8$ & $2$ & $4224$ & $\overline{\Gamma}$ & $\overline{Y}$ \tabularnewline
\hline 
$8.05$  & $5$ & $1$ & $1632$ & $\overline{X}$ & $\overline{M}$ \tabularnewline
\hline 
$6.72$  & $12$ & $2$ & $9344$ & $\overline{\Gamma}$ & $\overline{Y}$ \tabularnewline
\hline 
\end{tabular}
\caption{\label{tab:UC001}
Tilted unit cells $(001)$ summary.
First column: tilting angle, $\{n,m\}$ characteristic doublet for
a given unit cell; $N_{states}$ - number of states per unit cell
$V^{\prime}$. The two rightmost columns show the position of surface states in each case: 
TP means topologically protected, while TNP means topologically non-protected.}
\end{table}

\subsection{Tilted (111) surfaces}

In the case of tilted $(111)$ surfaces, the original (non-tilted)
unit cell is hexagonal and has the following unit vectors in Cartesian
coordinates: $\widetilde{a}_{1}=a\left(0,\frac{1}{2},-\frac{1}{2}\right),\;\widetilde{a}_{2}=a\left(-\frac{1}{2},0,\frac{1}{2}\right),\;\widetilde{a}_{3}=a\left(1,1,1\right)$.
However, this cell is not orthorhombic, therefore, we double and rotate
it by $45^{\circ}$ ($a_{1}=\widetilde{a}_{1}-\widetilde{a}_{2}$,
$a_{2}=\widetilde{a}_{1}+\widetilde{a}_{2}$) to end up with an orthorhombic
unit cell $V$ having, $a_{1}=a\left(\frac{1}{2},\frac{1}{2},-1\right),\;a_{2}=a\left(-\frac{1}{2},\frac{1}{2},0\right),\;a_{3}=a\left(1,1,1\right)$.
As above, we rotate $V$ around $a_{2}$, which becomes the new $b_{2}$.
We require that $b_{1}\perp b_{3}$, then, in general:
\begin{alignat*}{2}
b_{1} & =na_{1} & +m^{\prime}a_{3}\\
b_{3} & =-la_{1} & +ma_{3}.
\end{alignat*}
The condition of orthogonality imposes only one equation for four
unknowns. In order to minimize the size of $V^{\prime}$, we set $m^{\prime}=1$,
and since $\left|a_{1}\right|^{2}=\frac{3a^{2}}{2}$, while $\left|a_{3}\right|^{2}=3a^{2}$,
we arrive at the constraint 
\begin{equation}
nl=2m\label{eq:111_constraint}
\end{equation}
for the remaining three parameters. In Cartesian coordinates, the unit
vectors $\{b_{i}\}$ of $(111)$ tilted cell $V^{\prime}$ are expressed
as follows:
\begin{eqnarray*}
b_{1} & = & a\left(\begin{array}{c}
1+\frac{n}{2}\\
1+\frac{n}{2}\\
1-n
\end{array}\right),\;b_{2}=a\left(\begin{array}{c}
-\frac{1}{2}\\
\phantom{-}\frac{1}{2}\\
\phantom{-}0
\end{array}\right),\;b_{3}=a\left(\begin{array}{c}
m-\frac{l}{2}\\
m-\frac{l}{2}\\
m+l
\end{array}\right),
\end{eqnarray*}
with the constraint (\ref{eq:111_constraint}). This choice of basis ensures the orthorhombicity of
the unit cell $V^{\prime}$ and allows to build up a sequence of surfaces with the cleavage angle
gradually approaching zero. The cleavage angle in this setting depends only on $n$ as follows:
$\tan\vartheta=\frac{l}{\sqrt{2}m}=\frac{\sqrt{2}}{n}$, thanks to the constraint (\ref{eq:111_constraint}).

At $\vartheta=0$ (the pristine $(111)$ surface), there are three mirror planes: $(1\bar{1}0)$,
$(\bar{1}01)$ and $(0\bar{1}1)$.
It it easy to see that at finite $\vartheta$ the mirror planes $(\bar{1}01)$ and $(0\bar{1}1)$ are
lost, while the $(1\bar{1}0)$ is
preserved since the rotation axis $b_2$ is normal to it. Therefore, we conclude that the projections
of the TRS points on a tilted $(111)$ surface will have the topological protection if they lie 
within the $(1\bar{1}0)$ mirror plane. As we will show below, there are always such TSS along
with those without topological protection. We also derive the
general rule describing the location of TSS for given $n$, $m$ and $l$.

\begin{figure}
\includegraphics[angle=270,width=4.25cm]{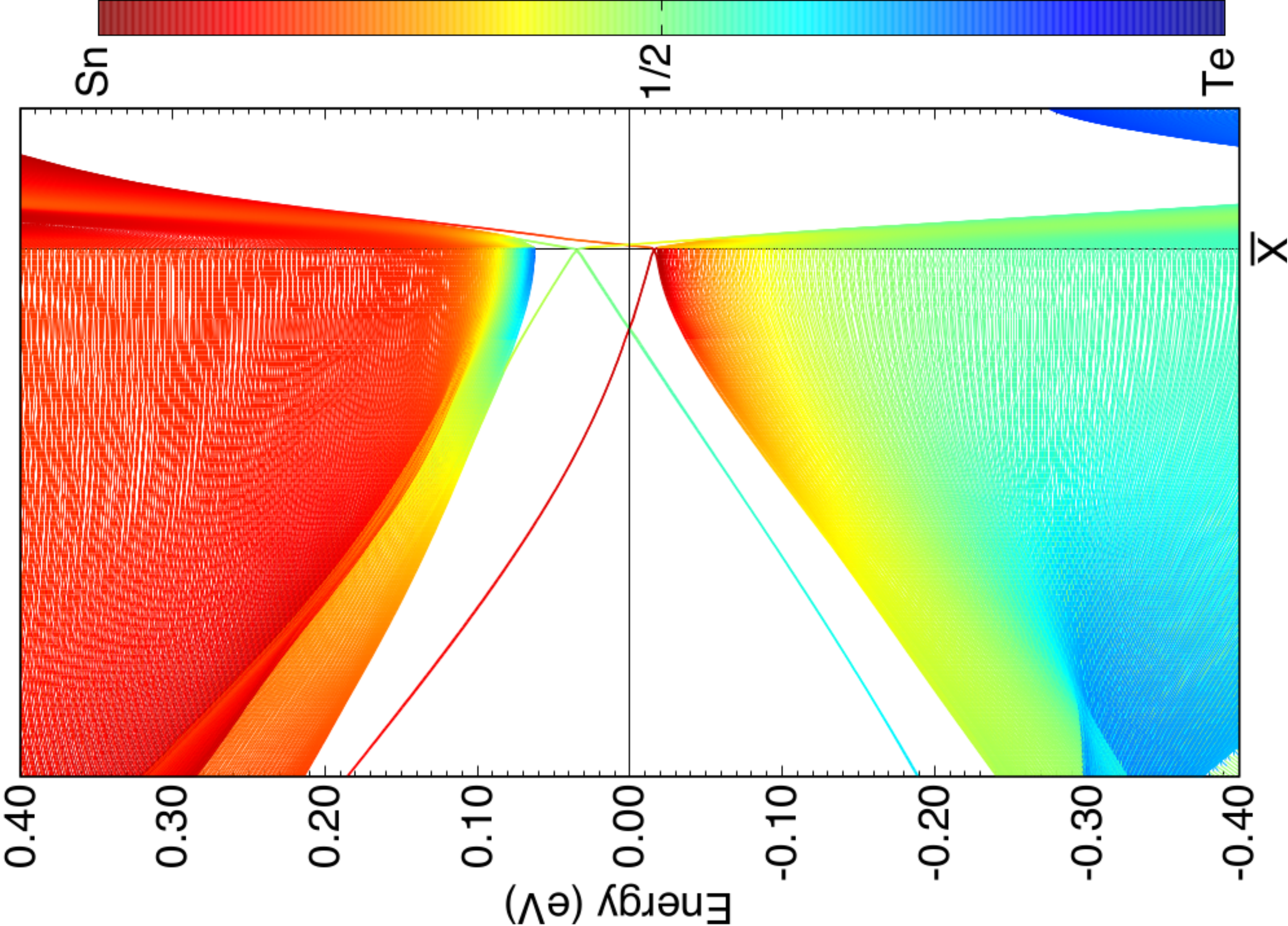}
\includegraphics[angle=270,width=4.25cm]{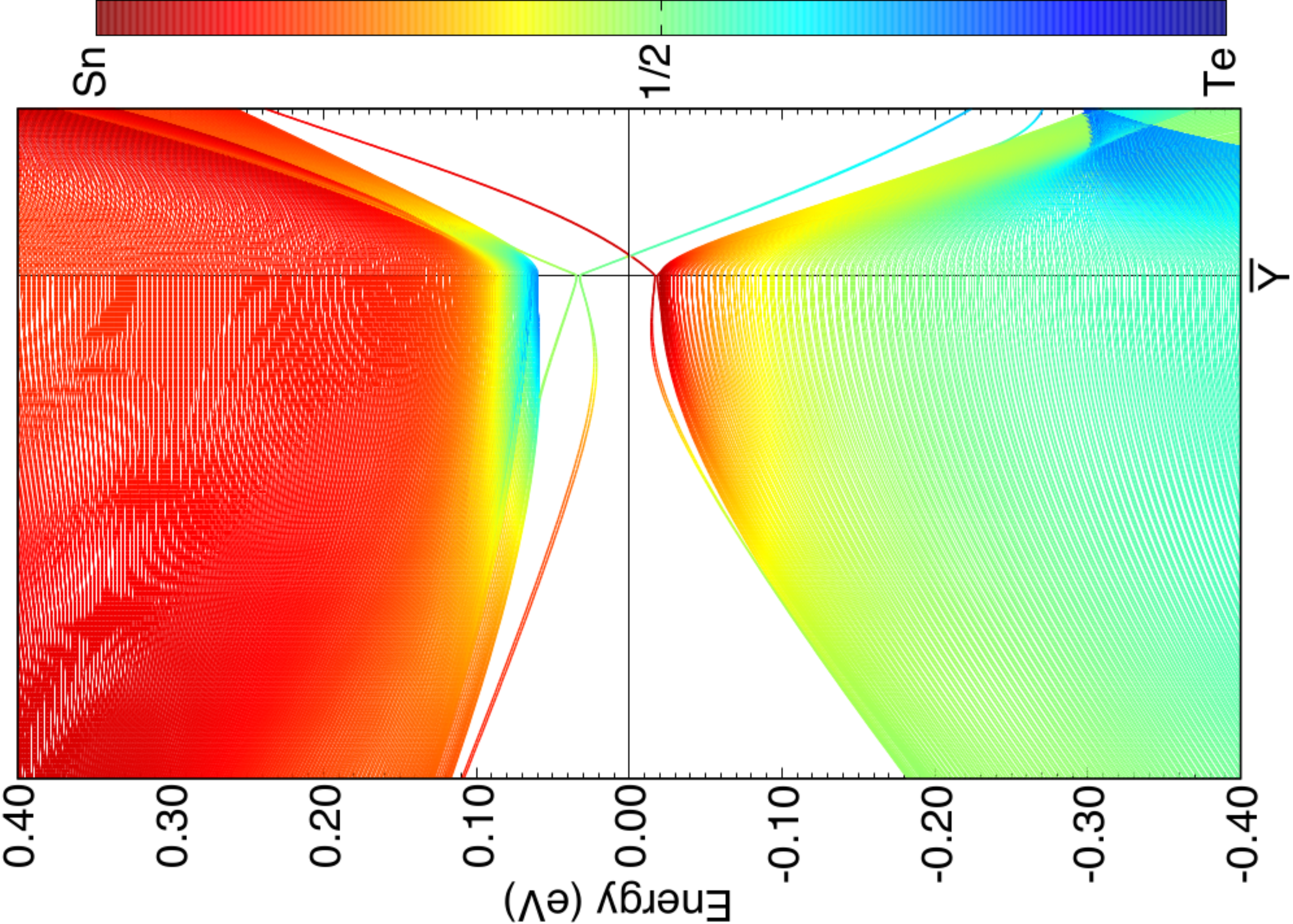}

\caption{\label{fig:TSS_111}(color online) 
   Surface band structure for a slab with tilted $(111)$ surface
   along the path $\overline{\Gamma}-\overline{X}-\overline{M}$ (left panel) and 
   $\overline{M}-\overline{Y}-\overline{\Gamma}$ (right panel). The
   line color reflects the orbital character of the bands: red - predominantly Sn, blue - predominantly Te.
   $n=1,$ $l=2$, $m=1$, which corresponds to the angle of
   $\alpha=54.74^{\circ}$, $110$ layers. For full details of the unit
   cell see Tab.\ref{tab:UC111}.
   Note the topological protection in the left panel and the absence thereof in
   the right one.
}
\end{figure}

\begin{figure*}
\begin{minipage}[t]{0.33\textwidth}%
\subfloat[$m=1$, $n=2$, $l=1$]{\includegraphics[viewport=120bp 90bp 900bp 910bp,clip,width=1\textwidth]{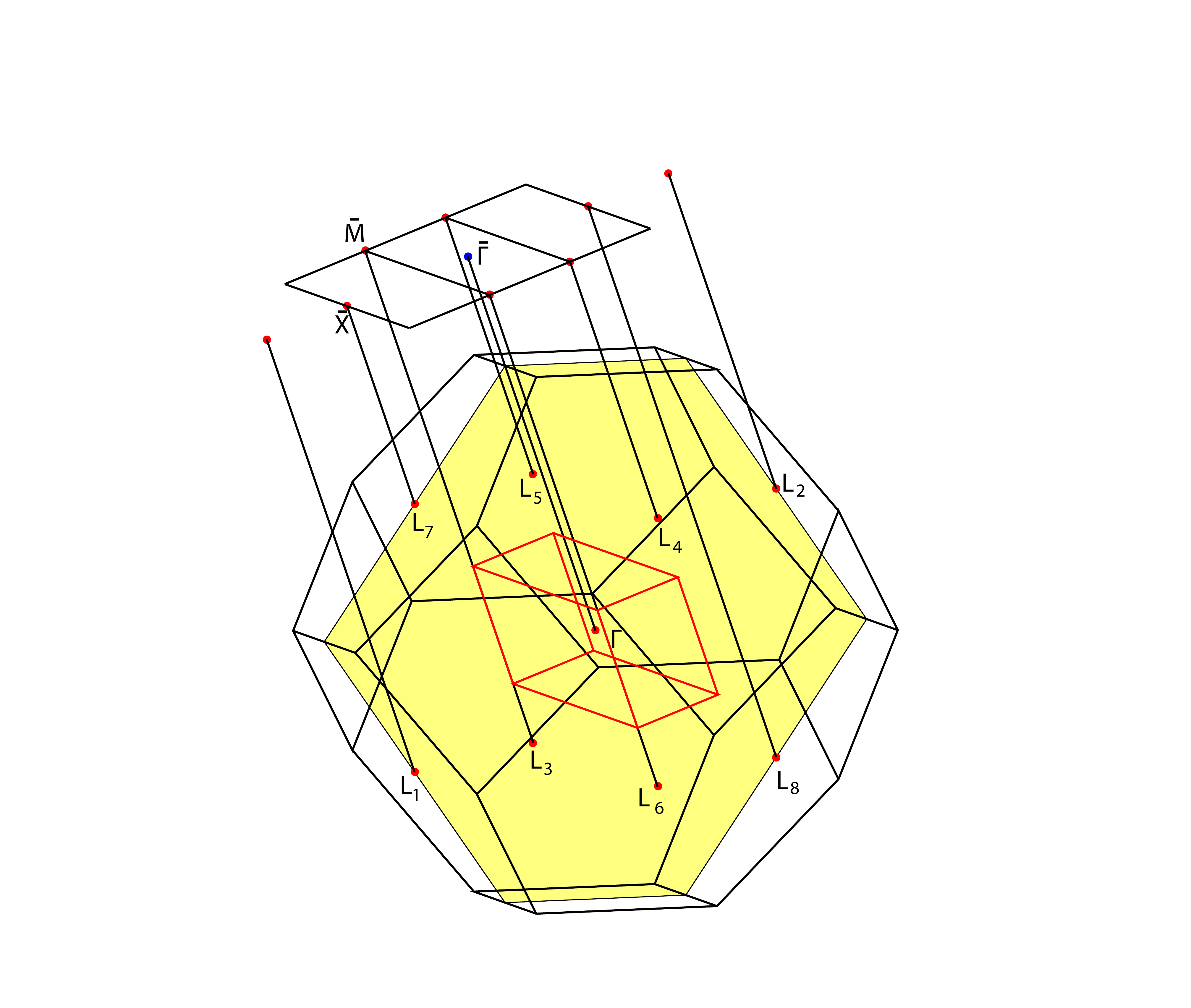}}%
\end{minipage}\hfill{}%
\begin{minipage}[t]{0.33\textwidth}%
\subfloat[$m=3$, $n=3$, $l=2$]{\includegraphics[viewport=120bp 90bp 900bp 910bp,clip,width=1\textwidth]{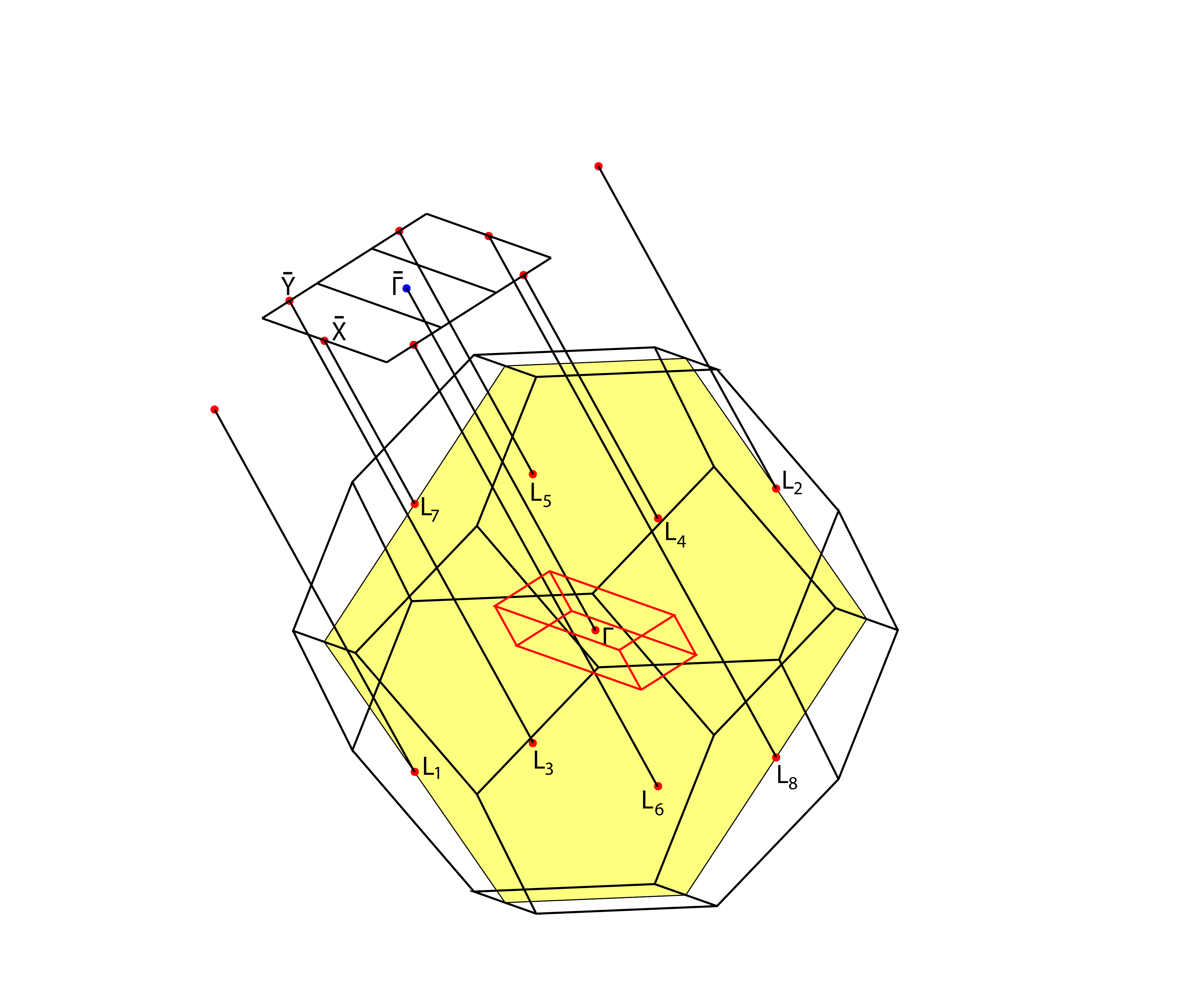}}%
\end{minipage}\hfill{}%
\begin{minipage}[t]{0.33\textwidth}%
\subfloat[$m=2$, $n=4$, $l=1$]{\includegraphics[viewport=120bp 90bp 900bp 910bp,clip,width=1\textwidth]{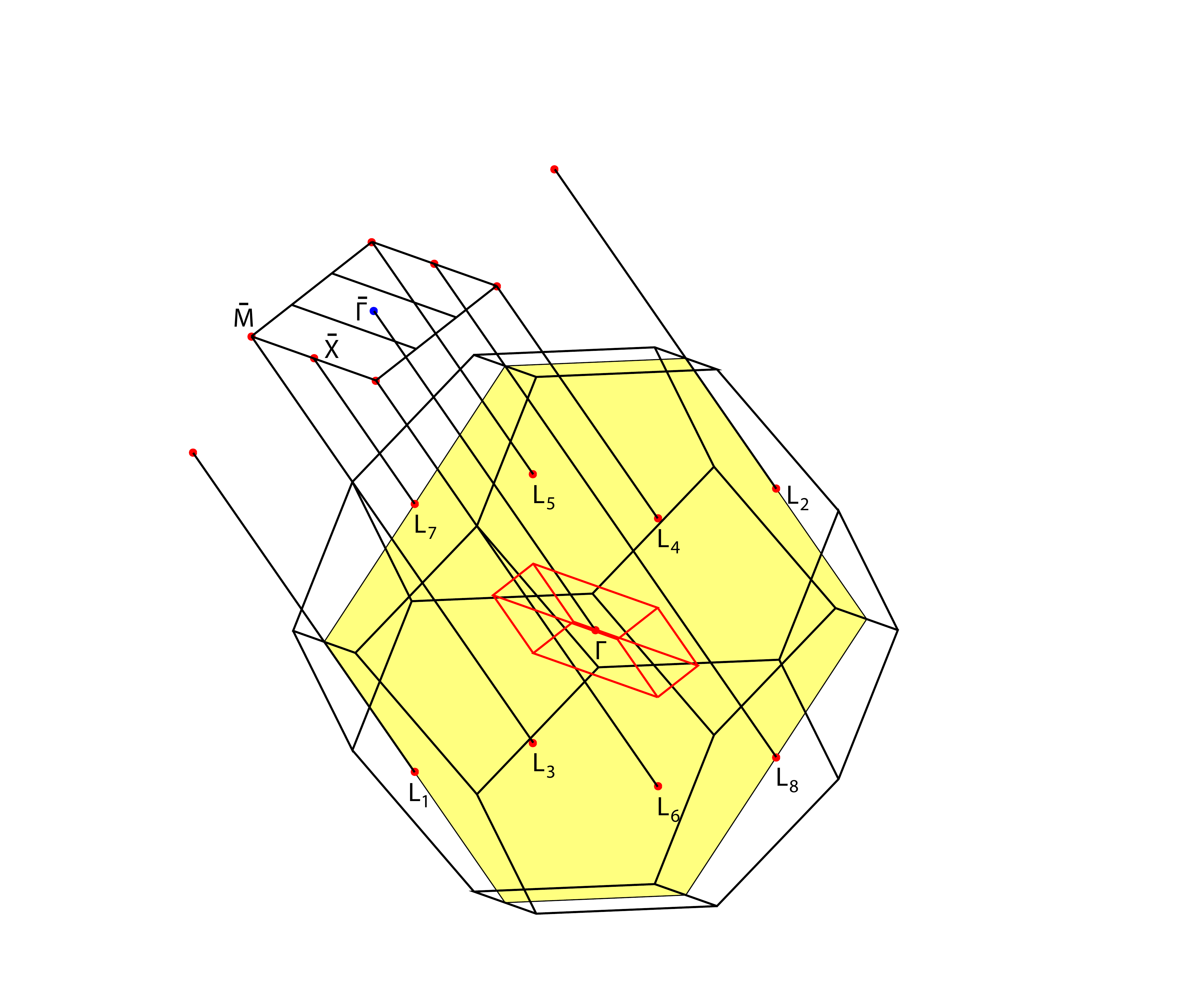}}%
\end{minipage}

\caption{\label{fig:BZ_111}(color online)
   The projections of $L$ TRS points of the rhombohedral Brillouin zone to a $(111)$ tilted
   surface. The corresponding folded BZs are also drawn in red. Three surface types are shown.
   Note the projection of $L$ points onto $\overline{X}$ and $\overline{Y}$ for (b), while
   projecting onto $\overline{X}$ and $\overline{M}$ for (a) and (c) cases. Shown in yellow is
   the $(1\bar{1}0)$ mirror plane.
}
\end{figure*}

The values of $\{n,l,m\}$ explored in the present work are listed in
Tab.\ref{tab:UC111}. An example of the slab dispersion with and without topological protection
in the case of $n=1$, $l=2$
and $m=1$ is depicted in Fig.~\ref{fig:TSS_111}. Once again there is a nice odd-even alternation
rule: for even $n$ the surface states appear at $\overline{M}=(\pi,\pi)$ and
$\overline{X}=(\pi,0)$, while for odd $n$ - at $\overline{Y}=(0,\pi)$ and $\overline{X}$, as
illustrated in Fig.~\ref{fig:BZ_111}.
Nevertheless, the topological protection is only ensured for the $\overline{X}$ points as they are
crossed by the $(1\bar{1}0)$ mirror plane, while $\overline{Y}$, and $\overline{M}$ are not
protected by the mirror symmetry.
When $n$ increases, the area of the first surface Brillouin
zone progressively diminishes, while the projections of the TRS points $L$ move towards higher
surface Brillouin zones, as in the case of the tilted $(001)$ surfaces. In the limit $n\to \infty$
the TRS projections turn to the usual picture seen on $(111)$ surfaces.
We notice that the alternation rule for TSS on $(111)$ surface does not depend on values of $l$ and
$m$ but only on $n$.

It is interesting to see how the lack of topological protection at $\overline{Y}$ on
Fig.~\ref{fig:TSS_111} leads to opening of a gap between the anion and cation topological states
on the left of $\overline{Y}$ point.
\begin{table}
\begin{tabular}{|c|c|c|c|c|c|c|}
\hline 
$\vartheta,\;^{\circ}$ & $n$ & $l$ & $m$ & $N_{states}$ & TP & TNP \tabularnewline
\hline 
\hline 
$54.74$ & $1$ & $2$ & $1$ & $288$ & $\overline{X}$ & $\overline{Y}$ \tabularnewline
\hline 
$35.26$ & $2$ & $1$ & $1$ & $288$ & $\overline{X}$ & $\overline{M}$ \tabularnewline
\hline 
$25.24$ & $3$ & $2$ & $3$ & $1056$& $\overline{X}$ & $\overline{Y}$ \tabularnewline
\hline 
$19.47$ & $4$ & $1$ & $2$ & $864$ & $\overline{X}$ & $\overline{M}$ \tabularnewline
\hline 
$15.79$ & $5$ & $2$ & $5$ & $2592$& $\overline{X}$ & $\overline{Y}$ \tabularnewline
\hline 
$13.26$ & $6$ & $1$ & $3$ & $1824$& $\overline{X}$ & $\overline{M}$ \tabularnewline
\hline 
$8.93$ & $9$ & $2$ & $9$ & $7968$ & $\overline{X}$ & $\overline{Y}$ \tabularnewline
\hline 
$6.72$ & $12$ & $1$ & $6$ & $7008$& $\overline{X}$ & $\overline{M}$ \tabularnewline
\hline 
\end{tabular}
\caption{\label{tab:UC111}Tilted unit cells $(111)$ summary. First column:
tilting angle, $\{n,l,m\}$ characteristic triplet for a given unit
cell; $N_{states}$ - number of states per unit cell $V^{\prime}$. The
two rightmost columns show the position of surface states in each case: 
TP means topologically protected, while TNP means topologically non-protected.}
\end{table}

\subsection{Tilted (110) surfaces}

\begin{table}[t]
\begin{tabular}{|c|c|c|c|c|c|}
\hline 
$\vartheta,\;^{\circ}$ & $n$ & $m$ & $N_{states}$ & TP & TNP \tabularnewline
\hline 
\hline 
$5.39$  & $1$ & $15$  & $7264$ & $\overline{X}$ & $\overline{M}$ \tabularnewline
\hline 
$8.05$  & $1$ & $10$  & $3264$ & $\overline{\Gamma}$ & $\overline{Y}$ \tabularnewline
\hline 
$13.26$ & $1$ & $6$  & $1216$  & $\overline{\Gamma}$ & $\overline{Y}$ \tabularnewline
\hline 
$22.00$ & $2$ & $7$  & $1824$  & $\overline{X}$ & $\overline{M}$ \tabularnewline
\hline 
$25.24$ & $3$ & $9$  & $3168$  & $\overline{X}$ & $\overline{M}$ \tabularnewline
\hline 
$29.50$ & $4$ & $10$ & $4224$  & $\overline{\Gamma}$ & $\overline{Y}$ \tabularnewline
\hline 
\end{tabular}
\caption{\label{tab:UC110}Tilted unit cells $(110)$ summary.
First column: tilting angle, $\{n,m\}$ characteristic doublet for
a given unit cell; $N_{states}$ - number of states per unit cell
$V^{\prime}$. The
two rightmost columns show the position of surface states in each case: 
TP means topologically protected, while TNP means topologically non-protected.}
\end{table}

The original cell $V$ in this case reads as:
$a_{1}=a\left(0,0,1\right),\;a_{2}=\frac{a}{2}(1,-1,0),\;a_{3}=\frac{a}{2}\left(1,1,0\right)$,
so that it has the pristine $a_{3}$ perpendicular to the $(110)$
surface. This unit cell can be viewed as a unit cell from Subsec.\ref{subsec:Tilted001}
with the vectors $a_{1}$ and $a_{3}$ interchanged. We have used
the following unit cell basis sets for this family of surface states
\begin{eqnarray*}
b_{1} & =a & \left(\begin{array}{l}
n\\
n\\
m
\end{array}\right),\quad b_{2}=\frac{a}{2}\left(\begin{array}{c}
\phantom{-}1\\
-1\\
\phantom{-}0
\end{array}\right),\quad b_{3}=a\left(\begin{array}{c}
\phantom{-}\frac{m}{2}\\
\phantom{-}\frac{m}{2}\\
-n
\end{array}\right)
\end{eqnarray*}
in Cartesian coordinates. It means that:
\begin{alignat*}{2}
   b_{1} & =\phantom{-}ma_{1} & +2na_{3}\\
b_{3} & =-na_{1} & +ma_{3}.
\end{alignat*}
This choice of basis ensures the orthorhombicity
of the unit cell and allows to build up a sequence of surfaces with
the cleavage angle gradually approaching zero. The cleavage angle
in this setting reads as follows: $\tan\vartheta=\sqrt{2}n/m$. It is easy to see that
$\vartheta_{(110)}=\frac{\pi}{2}-\vartheta_{(001)}$ as the unit cell in this case is the same as in
Sec.~\ref{subsec:Tilted001}. The pristine unit cell is recovered in the limit $m=1$, $n=0$.
The values of $\{n,m\}$ explored in the present work are listed in Tab.\ref{tab:UC110}.

\begin{figure*}
\begin{minipage}[t]{0.33\textwidth}%
\subfloat[$m=6$, $n=1$]{\includegraphics[viewport=120bp 90bp 900bp 910bp,clip,width=1\textwidth]{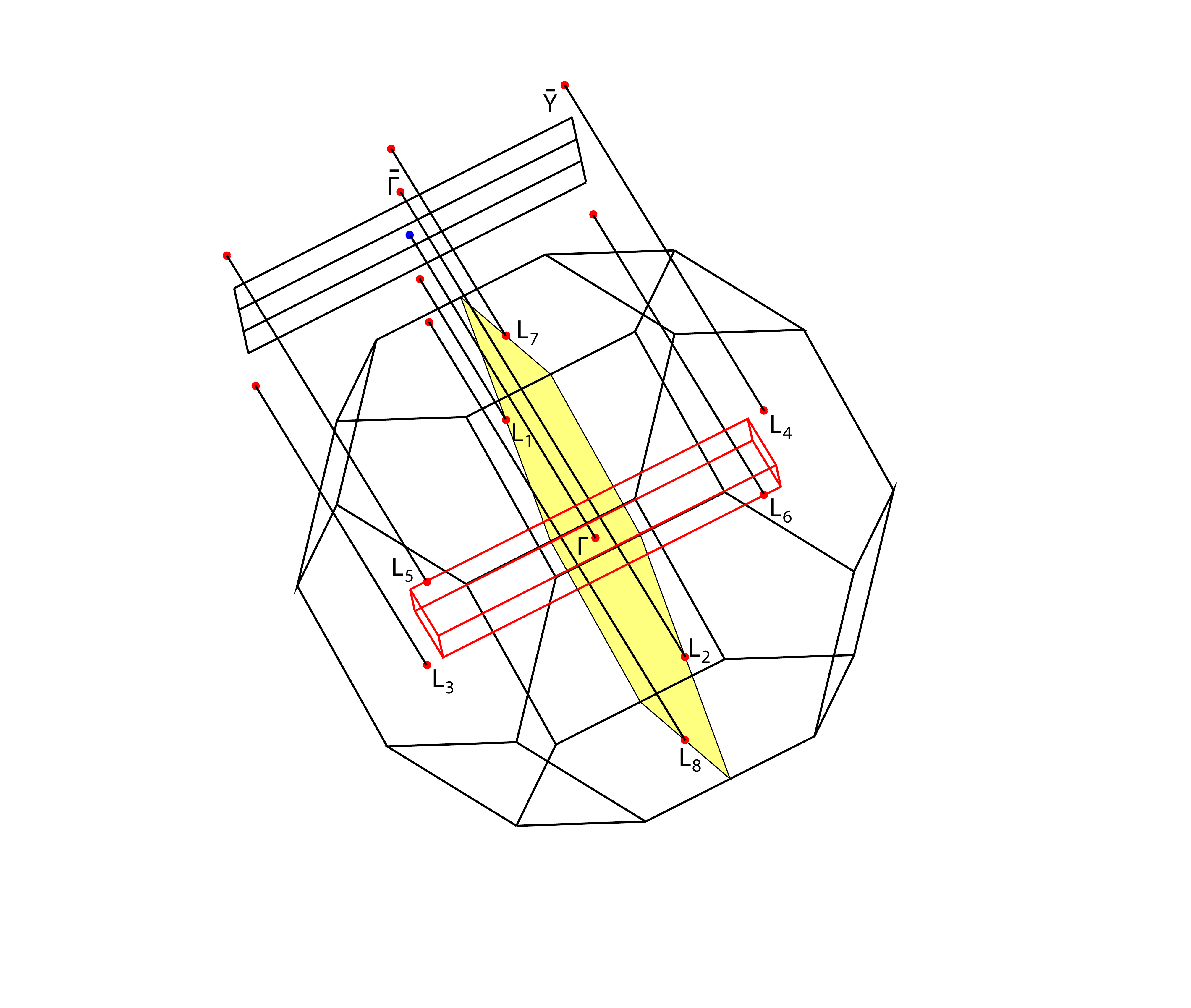}

}%
\end{minipage}\hfill{}%
\begin{minipage}[t]{0.33\textwidth}%
\subfloat[$m=7$, $n=2$]{\includegraphics[viewport=120bp 90bp 900bp 910bp,clip,width=1\textwidth]{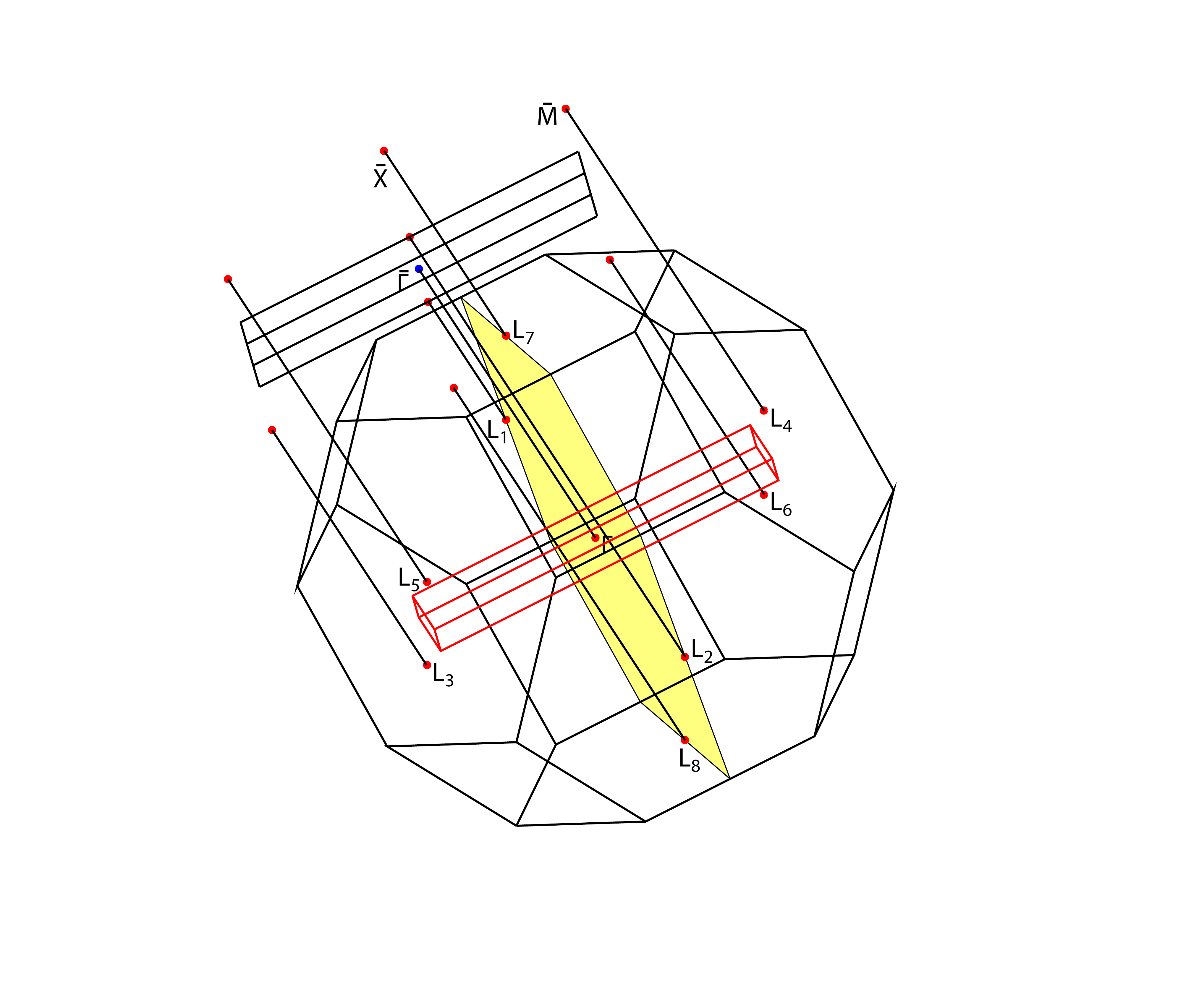}}%
\end{minipage}\hfill{}%
\begin{minipage}[t]{0.33\textwidth}%
\subfloat[$m=9$, $n=3$]{\includegraphics[viewport=120bp 90bp 900bp 910bp,clip,width=1\textwidth]{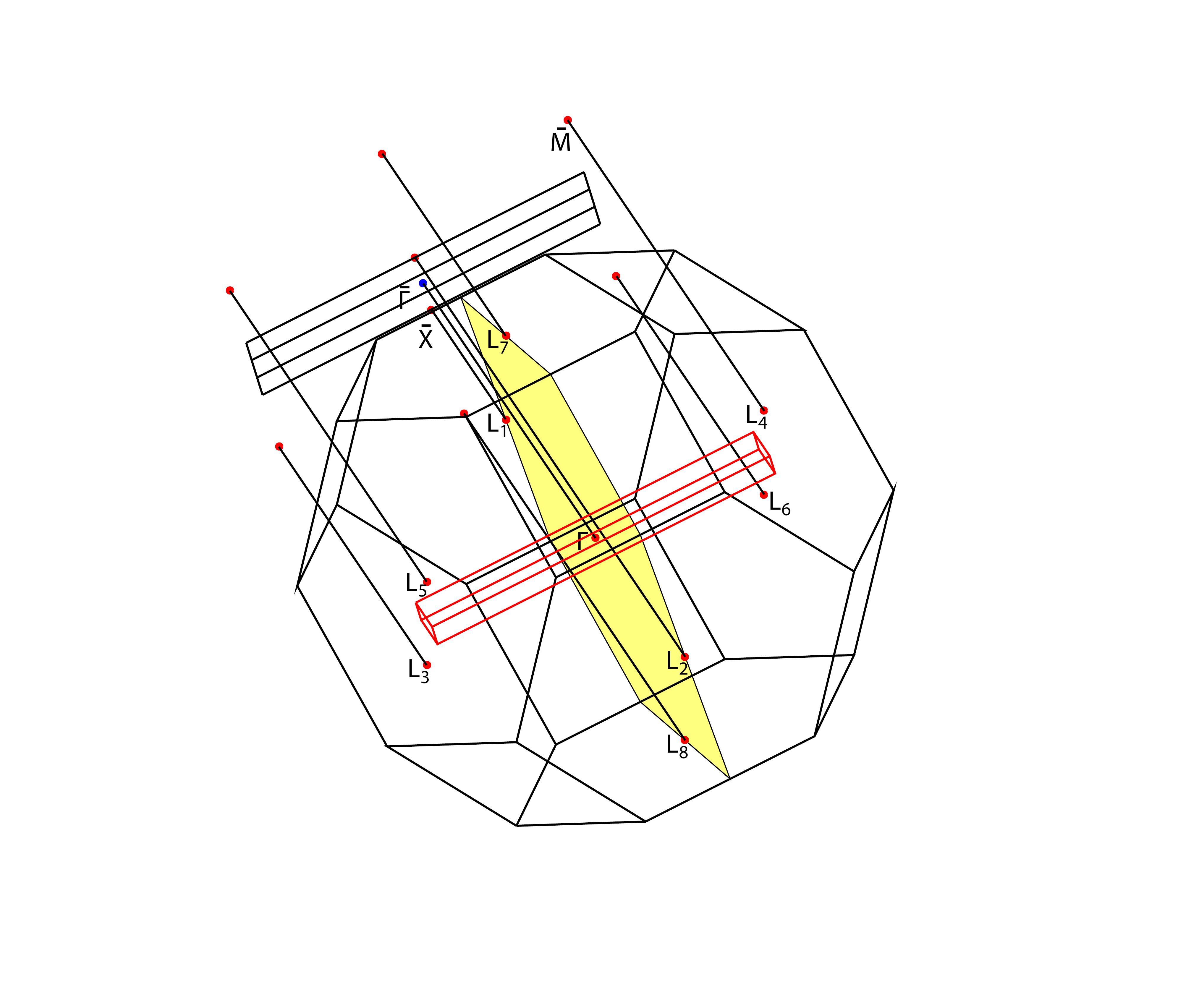}

}%
\end{minipage}

\caption{\label{fig:BZ_110}(color online)
   The projections of $L$ TRS points of the rhombohedral Brillouin zone to a $(110)$ tilted
   surface. The corresponding folded BZs are also drawn in red. Three surface types are shown.
   Note the projecting on $\overline{\Gamma}$ and $\overline{Y}$ for (a), while projecting on
   $\overline{M}$ and $\overline{X}$ for (b) and (c).
   Shown in yellow is the $(1\bar{1}0)$ mirror plane.}
\end{figure*}

\begin{figure}
\includegraphics[angle=270,width=4.25cm]{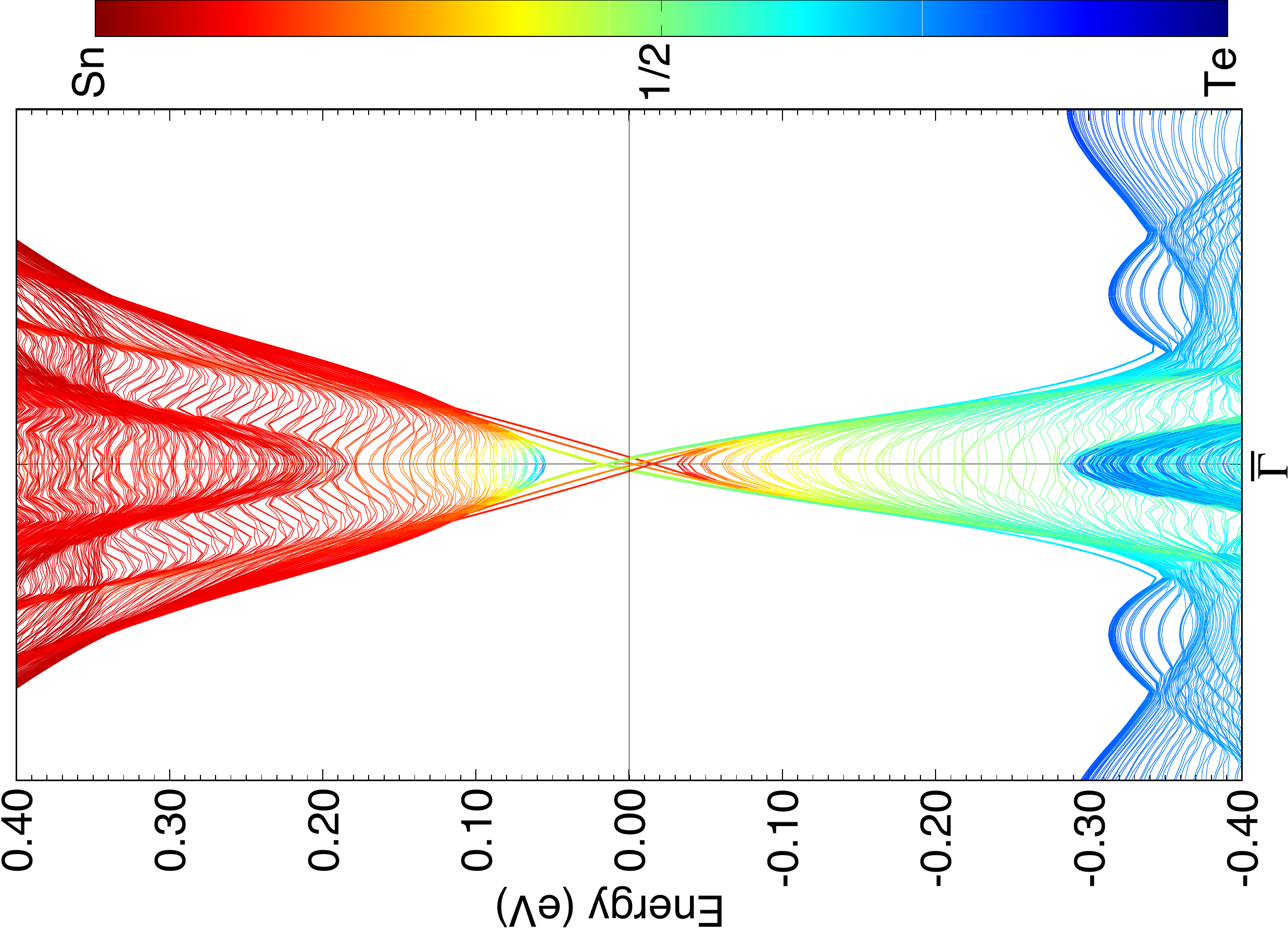}
\includegraphics[angle=270,width=4.25cm]{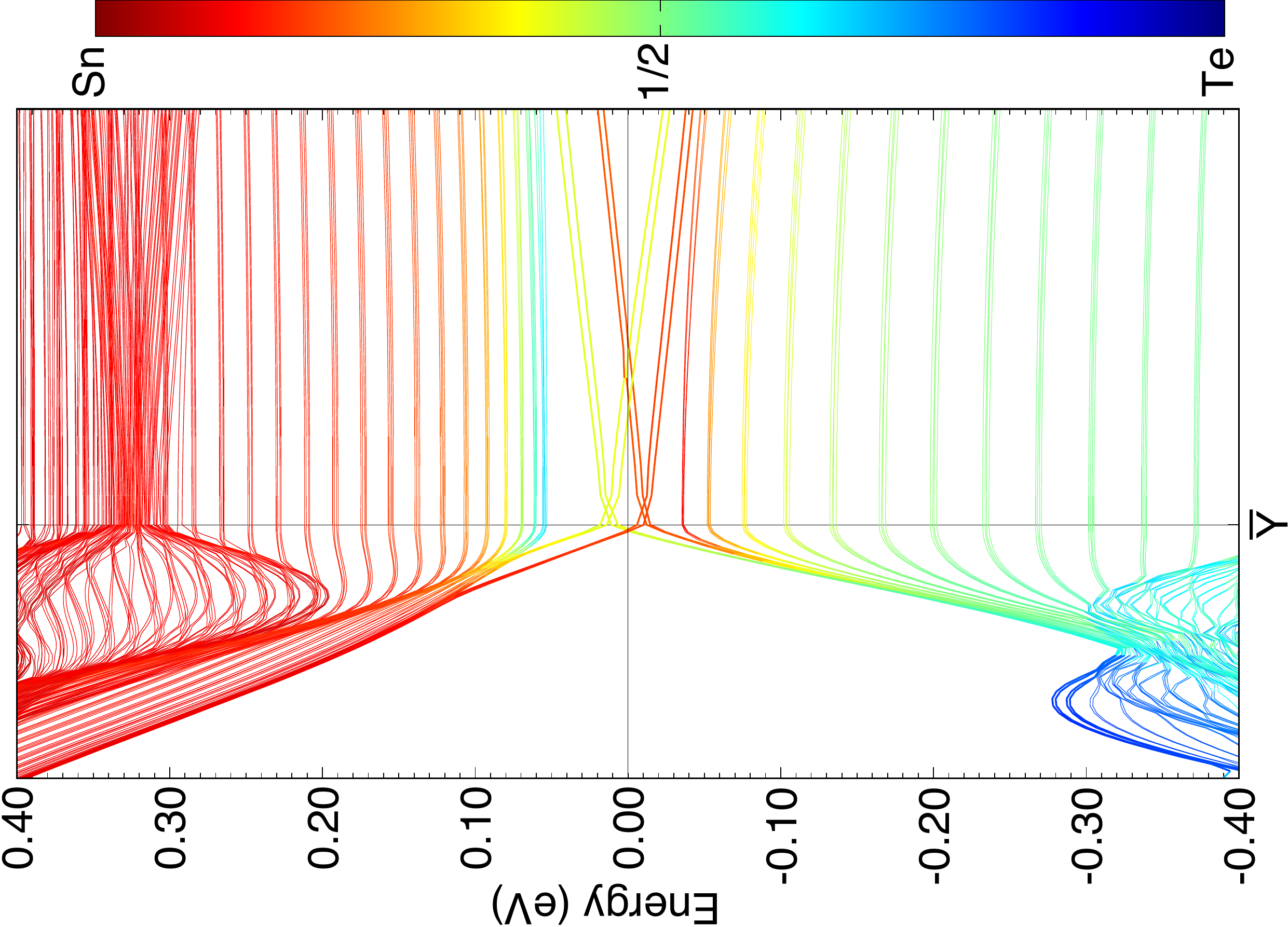}

\caption{\label{fig:TSS_110}(color online)
   Surface band structure for a slab with tilted $(110)$ surface
   along the path $\overline{Y}-\overline{\Gamma}-\overline{Y}$ zoomed around $\overline{\Gamma}$ (left panel) and 
   $\overline{\Gamma}-\overline{Y}-\overline{M}$ zoomed around $\overline{M}$ (right panel). The
   line color reflects the orbital character of the bands: red -predominantly Sn, blue -predominantly Te.
   $n=1$, $m=6$, which corresponds to the angle of $\alpha=13.26^{\circ}$.
   $18$ layers. For full details of the unit cell see Tab.\ref{tab:UC110}.
   Note the topological protection in the left panel and the absence thereof in
   the right one.
}
\end{figure}

Since $b_1$ and $b_2$ are the same as in Sec.~\ref{subsec:Tilted001}, the appearance
of the surface states follow the same rule outlined therein, namely independently on $n$, and
depending on parity of $m$. The difference with respect to the case of Sec.~\ref{subsec:Tilted001}
is that the limit of small $\vartheta$ now is realized at $m\gg n$ as opposed to $n\gg m$.
We report the 3D plots of such projections for $n=2,3,4$ in Fig.\ref{fig:BZ_110}.
It is interesting to note that each of the eight $L$ points of the
first primitive BZ is projected onto a different high symmetry point in different folded surface BZ.
Moreover, the projections of some of the $L$ points in the first bulk BZ belong to higher surface
BZs. 
As $n\to\infty$ (and $\vartheta\to0$),
the projections move more and more towards higher surface BZs.
On the other hand, as $n\to\infty$, the extension of the surface BZ
along $x$ direction tends to zero as well as the distance between
$\overline{Y}$ and $\overline{M}$.
We emphasize, that only $\overline{\Gamma}$ and $\overline{X}$ surface states have the topological
protection, while $\overline{M}$ and $\overline{Y}$ do not.
An example of the slab dispersion containing TSS in the case of $n=1$ and $m=6$ is depicted in Fig.~\ref{fig:TSS_110}.


\section{\label{sec:Conclusions}Conclusion}

In this manuscript we performed a systematic study of topological surface states on vicinal planes
with finite cleavage angle of the TCI tin telluride, by using the ab-initio derived effective
tight-binding model. We have set up TB calculations of slabs thick enough to observe the topological
surface states. We showed that the choice of vicinal plane has a direct consequence on the observed
topological states, and in particular on the number of the surface states and their positions.
We also discuss the limits $\vartheta\to 0$ and recovered the usual TSS in SnTe.
In particular, we found an
alternation rule, determining the position of the TSS based on the parity of one of the integers,
used to construct the folded unit cell. These integers, in turn, can be related to the cleavage
angle $\vartheta$ of the surface. In all the cases, the TSS appear at high-symmetry points
$\overline{\Gamma}$ and $\overline{X}$ as opposed to the limiting case $\vartheta=0$ and
$(001)$ where an exact coincidence of the projections of two $L$ points with a subsequent
hybridization and a shift off the high symmetry point occurs.
In the case of the vicinal plane surface states, only some of the states are topologically protected by
the mirror plane symmetry and are thus the topological surface states, contrarily to the case of the pristine surfaces where all the
TRS projections are topological surface states.
Our conclusions hold for arbitrary angles and orientations of the tilted surfaces.
The direct calculation of the mirror Chern numbers for the vicinal planes is in progress.

A singular feature of the TSS
projected onto the tilted vicinal planes consists in the fact that at small angles $\vartheta$,
the bulk TRS $L$ points from the first bulk Brillouin zone are projected not only onto the first surface 
Brillouin zone but also to the higher ones.

This work was supported by EPSRC (grants EP/R02992X/1). C.W. gratefully acknowledges the support of
NVIDIA Corporation with the donation of the Tesla K40 GPUs used for this research. For computational
resources, we were supported by the ARCHER UK National Supercomputing Service and the UK Materials
and Molecular Modelling Hub for computational resources (EPSRC Grant No. EP/ P020194/1).

\bibliographystyle{apsrev4-1}

%

\end{document}